%% file: main_arxiv.tex
\documentclass{article}

\usepackage{
  spconf,
  amsmath,
  amssymb,
  graphicx,
  booktabs,
  cite,
  url,
  tikz,
  siunitx,
  subcaption,
  hyperref
}

\newcommand{\sig}[1]{%
  \num[round-mode=figures,round-precision=2]{#1}%
}

\usetikzlibrary{arrows.meta,positioning,fit,calc}

\title{Ensemble-Based Perceptual Audio Quality Assessment\\
with Confidence Intervals}

\name{
  Pablo M. Delgado $^{1}$,
  Andreas Brendel $^{1}$\sthanks{Andreas Brendel has been supported by the Free
  State of Bavaria through the DSgenAI project.},
  Konstantin Schmidt $^{1}$,
  Jürgen Herre $^{2}$
}

\address{
  $^{1}$Fraunhofer Institute for Integrated Circuits IIS, Erlangen, Germany \\
  $^{2}$International Audio
  Laboratories Erlangen
  \sthanks{A joint institution of the Friedrich-Alexander Universität Erlangen-Nürnberg (FAU) and Fraunhofer IIS }, Germany
}

\begin{document}
\ninept
\maketitle

\begin{abstract}
Objective audio quality metrics typically provide point estimates, whereas
listening tests yield score distributions from which mean opinion scores
(MOS), confidence intervals (CIs), and significance decisions are derived.
We propose a lightweight intrusive metric that combines a PEAQ-style
perceptual front-end (ITU-R BS.1387) with a bagging ensemble of regressors.
Calibration with subjective data aligns ensemble outputs with listener scores.
The resulting item-dependent score distributions enable uncertainty
assessment, panel-size-matched CIs, and identification of less conclusive
predictions. Calibration improves agreement with subjective distributions and
CI coverage across all evaluated datasets while preserving MOS accuracy.
Using only 11 fixed PEAQ features, low-capacity regressors, and public
training data, the method performs comparably to more data-intensive
end-to-end approaches. Its output can support uncertainty-aware assessment
and target listening tests towards uncertain conditions. The distributions
also enable approximate pairwise comparisons, but not yet reliable
significance inference.
\end{abstract}


\section{Introduction}
\label{sec:intro}

Perceptual objective audio quality measurement (OAQM) systems are used to
benchmark codecs and enhancement algorithms, to monitor quality in
communication and streaming \cite{RixTaslp}, and as loss functions
\cite{zhen2020psychoacoustic} or evaluation metrics in machine-learning-based
generative audio \cite{FAD}. Classical OAQMs follow a two-stage design: a
signal-processing front-end implements a psychoacoustic model and extracts
features describing different degradation dimensions, and a compact
machine-learning (ML) stage, i.e., a regressor, maps these features to an
overall quality score using subjective mean opinion score (MOS) ratings, or similar
quantities, as training targets
\cite{PEAQ,Visqol3,delgado2024towards,POLQAcite}.

Subjective listening tests remain the gold standard for quality assessment. In standardised methods \cite{ITUP800,BS1116,MUSHRA}, results are commonly reported as MOS of Basic Audio Quality (BAQ) together with confidence intervals (CIs), reflecting uncertainty of the estimated mean, and with statistical comparisons between conditions. In contrast, most OAQMs perform only point estimation. However, it would be desirable to know the degree of uncertainty associated with a OAQM MOS prediction, whether the corresponding interval achieves adequate empirical coverage (i.e., the estimates' uncertainty matches the uncertainty achieved by subjective assessment methods), and whether an estimated quality difference would lead to the same decision as a listening test.

Beyond predicting a single MOS, the primary objective considered in this paper is to estimate the conditional distribution of subjective listener scores for a given reference--signal under test (SUT) pair. Such a distribution would allow a model to simulate listening-test panels, estimate MOS uncertainty, and assess the likelihood of significant differences between conditions. Listener-dependent models such as MBNet \cite{leng2021mbnet} and LDNet \cite{huang2022ldnet} use individual listener scores to model aspects of listener behaviour, but do not explicitly provide score distribution estimates. In contrast, the Generative Machine Listener (GML) \cite{jiang2023generative} explicitly models a conditional score density, typically by predicting the parameters of a Gaussian or logistic distribution, and samples simulated listener scores from this parametric density. These methods generally rely on deep end-to-end or self-supervised architectures and substantial training resources, while their representations are difficult to relate to established psychoacoustic models.

We propose a non-parametric alternative for estimating subjective score distributions based on an ensemble of
bootstrap-trained regressors \cite{dietterich2000ensemble,
lakshminarayanan2017simple} \footnote{Companion page is provided at: \url{http://audiolabs-erlangen.de/resources/2027-ICASSP-ELQA} (in preparation after review)}. For each SUT, the ensemble outputs form a
Monte Carlo (MC) sample distribution of MOS predictions rather than the parameters of a prespecified density. This approach is motivated by the assumption that shuffling the MOS-labelled training material produces different plausible learned quality mappings, analogous to listeners with different experience or preferences. We refer to
this assumption as the artificial-listener hypothesis. We test whether a simple calibration of the ensemble output based on subjective data is sufficient for ensemble scores to act as proxies for human listener scores. We evaluate this using distributional measures and listening-test-style decisions.

\begin{figure*}[t]
  \centering
  \resizebox{0.91\textwidth}{!}{%
    \input{method_diagram.tex}%
  }
  \caption{Proposed system. A shared PEAQ-style front-end feeds
bootstrap-trained regressors. Affine mean and global width calibration
convert their outputs into candidate artificial-listener scores, from which
MOS and panel-matched CIs are estimated. Solid and dashed arrows denote
inference and training or calibration, respectively.}
  \label{fig:blockdiagram}
\end{figure*}
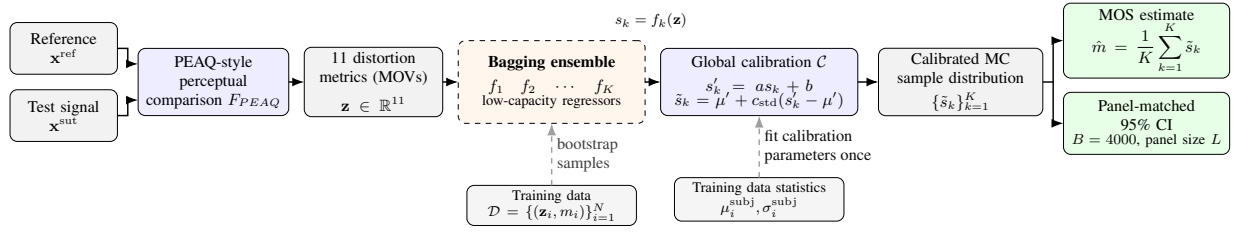

\section{Method}
\label{sec:method}

Fig.~\ref{fig:blockdiagram} illustrates the proposed approach. For each reference signal and SUT comparison, distortion metrics are computed relative to its reference and mapped to MOS by a regressor ensemble.

\subsection{System Overview and Perceptual Front-End}

Let $\mathbf{x}^{\mathrm{ref}},\mathbf{x}^{\mathrm{sut}}\in\mathbb{R}^{T\times C}$ denote the
reference and SUT digital audio signals with $T$ time samples and $C$ channels. A non-trainable intrusive PEAQ-style
front-end\footnote{The PEAQ front-end implementation is available at:
\url{https://www.mmsp.ece.mcgill.ca/Documents/Software/}.}
extracts a feature vector
$\mathbf{z}~=F_{PEAQ}(\mathbf{x}^{\mathrm{ref}},\mathbf{x}^{\mathrm{sut}})\in~\mathbb{R}^{J}$ shared by all regressors.
We use the $J=11$ model output variables (MOVs) of the basic PEAQ model
\cite{PEAQ,ThiedePEAQ}. They describe perceptually relevant degradations,
including bandwidth limitation, noise, and temporal or spectral
distortions, and are computed once for each signal pair.

The feature vector is passed to $K$ low-capacity regressors, producing
$s_k=f_k(\mathbf{z})\in\mathbb{R}$ for $k=1,\ldots,K$.
The regressors may use one or more multivariate regression algorithms and
are trained on different bootstrap samples.
Before calibration, the variation among their outputs reflects the
sensitivity of the learned MOS mapping to the bootstrap-resampled training
data. After calibration, the resulting score distribution is expected to more closely resemble subjective score distributions.
A global calibration $\mathcal{C}:\mathbb{R}^K\rightarrow\mathbb{R}^K$ converts the raw member scores into
calibrated scores $\tilde{s}_k$:
\begin{equation}
(\mathbf{x}^{\mathrm{ref}},\mathbf{x}^{\mathrm{sut}})
\xrightarrow{\;F_{PEAQ}\;}
\mathbf{z}
\xrightarrow{\;\{f_k\}_{k=1}^{K}\;}
\{s_k\}_{k=1}^{K}
\xrightarrow{\;\mathcal{C}\;}
\{\tilde{s}_k\}_{k=1}^{K}.
\end{equation}

\subsection{Ensemble Training and Monte Carlo Resampling}
\label{subsec:ensemble_training}

For each training stimulus $i=1,\ldots,N$, the front-end produces a feature vector
$\mathbf{z}_i$. Given its corresponding $L_i$ listener scores
$y_{i\ell}\in[0,100]$ \footnote{This particular scale range corresponds to the subjective listening test method of choice for the ground-truth data.}, $\ell\in\{1,\dots,L_i\}$, the subjective MOS representing the regression target is
$m_i=L_i^{-1}\sum_{\ell=1}^{L_i}y_{i\ell}$. Here, $m_i$ denotes the observed subjective MOS for stimulus $i$, viewed as an estimate of the corresponding population mean.
The regression training data are therefore
$\mathcal{D}=\{(\mathbf{z}_i,m_i)\}_{i=1}^{N}$.

For each ensemble member $k$, we draw a bootstrap sample (MC resample)
$\mathcal{D}^{(k)}$ of size $N$ from $\mathcal{D}$ with replacement and fit each regressor by squared-error minimisation.

For each stimulus, the mean $\mu^{\mathrm{mc}}$ and standard deviation $\sigma^{\mathrm{mc}}$ of the ensemble output scores could be used as a point estimate of subjective MOS and a measure of score variability induced by the training-data partitions, respectively\cite{DelgadoPEAQ}. However, this distribution does not necessarily reflect subjective-score distributions accurately. Raw ensemble means can be biased or incorrectly scaled relative to subjective scores, and raw ensemble widths need not match subjective score variability. We therefore estimate one global mean-and-width calibration using the training database as ground truth to be applied to the raw score distribution.

\subsection{Global Mean-and-Width Calibration}
\label{subsec:affine}
In order to align location and scale parameters of the MC distributions to the distribution of subjective scores, we first apply an affine transformation with global parameters $a$ and $b$ to
each ensemble-member score:
$
s'_k=as_k+b.
$
This transformation maps the ensemble mean according to $\mu'=a\mu+b$ and
scales the standard deviation of the ensemble-member scores by $|a|$. The parameters are fitted by least squares:
\begin{equation}
(a,b)
=
\arg\min_{a,b}
\sum_i
\left[
m_i
-
\left(a\mu_i^{\mathrm{mc}}+b\right)
\right]^2.
\label{eq:affine_mean}
\end{equation}
To align the scale parameters of the ensemble-score distribution with the observed
listener-score spread, the deviations of the affinely mapped member scores are
rescaled
\begin{equation}
\tilde{s}_k
=
\mu'
+
c_{\mathrm{std}}\left(s'_k-\mu'\right),
\qquad k=1,\ldots,K,
\label{eq:member_calibration}
\end{equation}
where $\mu'=K^{-1}\sum_k s'_k$. This operation preserves the affine-calibrated arithmetic mean $\mu'$ while
scaling the standard deviation of the member scores by $c_{\mathrm{std}}$. We estimate this global width factor as
\begin{equation}
c_{\mathrm{std}}
=
\operatorname{median}_{i}
\frac{\sigma_i^{\mathrm{subj}}}
     {|a|\sigma_i^{\mathrm{mc}}},
\label{eq:spread_calibration}
\end{equation}
where $\sigma_i^{\mathrm{subj}}$ is the sample standard deviation of the
individual listener scores for stimulus $i$. We assume that both subjective
and ensemble variances are finite and strictly positive, as is the case for
the data used in the experiments.

In the proposed two-stage transformation  Eq.~\eqref{eq:affine_mean} aligns the
predicted mean, while Eq.~\eqref{eq:spread_calibration} sets one global
spread factor such that the median ratio of subjective to calibrated ensemble
standard deviation across the calibration stimuli equals one. From the model outputs $\tilde{s}_k$ forming an estimated empirical score distribution, point estimates
$
\hat{m}
=
\frac{1}{K}\sum_{k=1}^{K}\tilde{s}_k
$
and confidence intervals can be calculated.

\subsection{Panel-Size-Matched CI Estimation}
\label{subsec:ci}

To compare the calibrated model distribution with a listening test
containing $L$ listeners, we simulate $B=4000$ pseudo-panels of the same
size. For the $b$th bootstrap draw, calibrated member scores are sampled with
replacement and averaged:

\begin{equation}
\bar{s}^{(b)}(L)
=
\frac{1}{L}
\sum_{\ell=1}^{L}
\tilde{s}_{I_{b\ell}},
\qquad
I_{b\ell}\sim\mathcal{U}\{1,\ldots,K\}.
\label{eq:pseudo_panel}
\end{equation}
Here, $\mathcal{U}\{1,\ldots,K\}$ denotes the discrete uniform distribution over the $K$ ensemble-member indices. The nominal 95\% panel-matched ensemble CI is given by the empirical
quantiles of the resulting pseudo-panel means \footnote{\(Q_{\bar{s}_i(L)}(p)\) denotes the empirical \(p\)-quantile of the Monte Carlo distribution of panel-mean scores \(\bar{s}_i(L)\) for stimulus \(i\) and panel size \(L\).}:

\begin{equation}
\mathrm{CI}^{\mathrm{mc}}_{i,95}(L)
=
\left[
Q_{\bar{s}_i(L)}(\alpha),
Q_{\bar{s}_i(L)}(1-\alpha)
\right].
\label{eq:mc_ci}
\end{equation}
with $\alpha=0.025$. Under the artificial-listener hypothesis, this interval estimates the
variation that would be obtained by repeatedly drawing panels of $L$
calibrated model listeners.

\section{Experiments}
\label{sec:experiments}

To assess the artificial-listener hypothesis, we evaluate three increasingly
demanding implications: empirical MOS coverage, agreement of marginal score
distributions, and agreement of paired significance decisions.

\subsection{Training and Evaluation Databases}

The ensemble and global calibration are trained on SEBASS-DB
\cite{SEBASSDB}, a public meta-dataset of MUSHRA \cite{MUSHRA} listening tests
on the BAQ of blindly separated audio sources. SEBASS-DB
combines SASSEC, SiSEC08, SAOC-DB, and SiSEC18, providing more than 11\,000
ratings for over 900 separated signals. It was selected for its public
availability.

For cross-domain evaluation in the audio-coding domain, we use the Open
Dataset of Audio Quality (ODAQ) \cite{ODAQ} and the Unified Speech and Audio
Codec (USAC) verification sets \cite{USACdatabase}. Preliminary training on
ODAQ was less effective, possibly because its emphasis on isolated coding
artefacts provides fewer examples of the artefact interactions encountered in
practice and needed by the regressors. Both evaluation sets contain stereo
reference excerpts comprising music, speech, dialogue, and movie-like
soundtracks, processed using different degradations (\emph{conditions}). All
datasets provide reference signals, degraded signals, and listener scores on
a 0 (\emph{bad}) to 100 (\emph{excellent}) quality scale.

\begin{table*}[t]
  \centering
  \caption{Results for MOS accuracy, uncertainty, and decisions. R/C:
raw/calibrated ensemble. PCC entries are R/C/2f/DeePAQ; RMSE entries
are R/C/2f; all others are R/C. DeePAQ PCC values are from \cite{DeePAQ}.
Pair counts are all/significant; ODAQ pools six experiments. Coverage is
below \(0.95\), so higher is better. Bold marks the best value per dataset
and metric.}
  \label{tab:combined_results}
  \footnotesize
  \renewcommand{\arraystretch}{1.05}
  \setlength{\tabcolsep}{3pt}

  \begin{tabular*}{\textwidth}{@{\extracolsep{\fill}}lcrrrrrr@{}}
    \toprule
    & &
    \multicolumn{2}{c}{Scalar MOS Accuracy (R/C/2f/DeePAQ)} &
    \multicolumn{2}{c}{Uncertainty (R/C)} &
    \multicolumn{2}{c}{Decision Agreement (R/C)} \\
    \cmidrule(lr){3-4}
    \cmidrule(lr){5-6}
    \cmidrule(lr){7-8}

    Dataset &
    \shortstack{Pairs\\all/sig.} &
    PCC \(\uparrow\) &
    RMSE \(\downarrow\) &
    \(\widehat{C}_{95}\uparrow\) &
    \(\overline{D}_{\mathrm{KS}}\downarrow\) &
    \(R_{\mathrm{sig}}\uparrow\) &
    FPR \(\downarrow\) \\
    \midrule

    SAOC-DB & 153\,/\,16 &
    \textbf{\sig{0.93}}\,/\,\textbf{\sig{0.93}}\,/\,\sig{0.81}\,/\,\textbf{\sig{0.934}} &
    \sig{5.21}\,/\,\textbf{\sig{5.17}}\,/\,\sig{8.06} &
    \sig{0.30}\,/\,\textbf{\sig{0.82}} &
    \sig{0.50}\,/\,\textbf{\sig{0.33}} &
    \textbf{\sig{0.94}}\,/\,\sig{0.69} &
    \sig{0.84}\,/\,\textbf{\sig{0.60}} \\

    SASSEC & 55\,/\,40 &
    \textbf{\sig{0.97}}\,/\,\textbf{\sig{0.97}}\,/\,\sig{0.90}\,/\,\sig{0.920} &
    \sig{4.39}\,/\,\textbf{\sig{4.38}}\,/\,\sig{7.42} &
    \sig{0.31}\,/\,\textbf{\sig{0.75}} &
    \sig{0.46}\,/\,\textbf{\sig{0.28}} &
    \textbf{\sig{1.00}}\,/\,\sig{0.85} &
    \sig{0.80}\,/\,\textbf{\sig{0.27}} \\

    SiSEC08 & 36\,/\,27 &
    \textbf{\sig{0.96}}\,/\,\textbf{\sig{0.96}}\,/\,\sig{0.91}\,/\,\sig{0.948} &
    \sig{4.63}\,/\,\textbf{\sig{4.62}}\,/\,\sig{6.72} &
    \sig{0.28}\,/\,\textbf{\sig{0.69}} &
    \sig{0.47}\,/\,\textbf{\sig{0.30}} &
    \textbf{\sig{1.00}}\,/\,\sig{0.89} &
    \sig{0.78}\,/\,\textbf{\sig{0.22}} \\

    ODAQ & 87\,/\,82 &
    \sig{0.85}\,/\,\sig{0.83}\,/\,\sig{0.87}\,/\,\textbf{\sig{0.916}} &
    \sig{11.40}\,/\,\sig{11.85}\,/\,\textbf{\sig{10.01}} &
    \sig{0.34}\,/\,\textbf{\sig{0.47}} &
    \sig{0.44}\,/\,\textbf{\sig{0.41}} &
    \textbf{\sig{0.57}}\,/\,\sig{0.29} &
    \sig{0.40}\,/\,\textbf{\sig{0.20}} \\

    USAC1 & 36\,/\,34 &
    \textbf{\sig{0.94}}\,/\,\sig{0.93}\,/\,\sig{0.86}\,/\,\sig{0.900} &
    \textbf{\sig{5.20}}\,/\,\sig{5.67}\,/\,\sig{7.89} &
    \sig{0.14}\,/\,\textbf{\sig{0.33}} &
    \sig{0.43}\,/\,\textbf{\sig{0.36}} &
    \textbf{\sig{1.00}}\,/\,\sig{0.94} &
    \textbf{\sig{0.50}}\,/\,\textbf{\sig{0.50}} \\

    USAC2 & 21\,/\,20 &
    \sig{0.81}\,/\,\sig{0.80}\,/\,\sig{0.64}\,/\,\textbf{\sig{0.875}} &
    \textbf{\sig{8.11}}\,/\,\sig{8.39}\,/\,\sig{11.17} &
    \sig{0.20}\,/\,\textbf{\sig{0.43}} &
    \sig{0.41}\,/\,\textbf{\sig{0.34}} &
    \textbf{\sig{0.90}}\,/\,\sig{0.85} &
    \sig{1.00}\,/\,\textbf{\sig{0.00}} \\

    USAC3 & 28\,/\,27 &
    \sig{0.78}\,/\,\sig{0.77}\,/\,\sig{0.82}\,/\,\textbf{\sig{0.928}} &
    \sig{9.58}\,/\,\sig{9.52}\,/\,\textbf{\sig{8.54}} &
    \sig{0.22}\,/\,\textbf{\sig{0.34}} &
    \sig{0.42}\,/\,\textbf{\sig{0.35}} &
    \textbf{\sig{0.74}}\,/\,\sig{0.70} &
    \sig{1.00}\,/\,\textbf{\sig{0.00}} \\

    \bottomrule
  \end{tabular*}
\end{table*}

\subsection{Base Learner Selection}

Multivariate Adaptive Regression Splines (MARS) were selected as the base
learner for their accuracy--complexity trade-off and smooth behaviour
\cite{DelgadoPEAQ}. Learner selection and hyperparameter tuning used repeated
50/50 holdout resampling on SEBASS-DB; the final ensemble members were fitted
on bootstrap resamples as described in Sec.~\ref{subsec:ensemble_training}.
For these experiments, \(K=50\) base regressors provided a practical
accuracy--complexity trade-off, yielding sufficiently stable predictions
without unnecessary computational cost. Experiments used Python 3.10,
\texttt{scikit-learn} 1.7.2, and
\texttt{py-earth2} 0.2.0.\footnote{\url{https://scikit-learn.org/};
\url{https://pypi.org/project/py-earth2/}.}

\subsection{Operational Evaluation}
\label{subsec:operational_eval}

All analyses are performed separately by database or listening-test
experiment for raw (R) and calibrated (C) outputs.

\textbf{Distributional Agreement:} As a compact measure of listener-score distribution agreement, we compute
the two-sample Kolmogorov\nobreakdash--Smirnov (KS) statistic separately for every
stimulus $i$.
The KS statistic is defined as

\begin{equation}
D_{\mathrm{KS},i}
=
\sup_x
\left|
F^{\mathrm{subj}}_i(x)
-
F^{\mathrm{mc}}_i(x)
\right|,
\end{equation}
where $F^{\mathrm{subj}}_i$ and $F^{\mathrm{mc}}_i$ are the empirical
cumulative distribution functions (CDFs) of the subjective and ensemble scores for
stimulus $i$. A smaller value indicates
less worst-case mismatch between CDFs and reduces the risk of locally unreliable proxies.

For each database, we report the mean per-stimulus KS statistic,
$
\overline{D}_{\mathrm{KS}}
=
\frac{1}{N}
\sum_{i=1}^{N}D_{\mathrm{KS},i}.
$

\textbf{MOS Accuracy and Empirical CI Coverage: }
For conventional OAQM evaluation, we report the point-estimate MOS comparison between subjective and objective means \cite{EvalObjective}. Model outputs are
clipped to this range, except in the significant-difference analysis, where
clipping is omitted to avoid boundary-induced effects. The Pearson correlation coefficient (PCC) and the root mean squared error (RMSE) are computed from the per-item subjective and ensemble means. We also report DeePAQ and the 2f metric \cite{2fKastner} as a comparable state-of-the-art baseline, as 2f's regressor was likewise fitted on SEBASS-DB. Results for other state-of-the-art metrics on these datasets are provided in \cite{DeePAQ}.

For each item, the objective CI is the panel-matched bootstrap interval in~\eqref{eq:mc_ci}, with $L=L_i$. Its empirical coverage is

\begin{equation}
\widehat{C}_{95}
=
\frac{1}{N}
\sum_{i=1}^{N}
\mathbb{I}
\left[
m_i
\in
\mathrm{CI}^{\mathrm{mc}}_{i,95}(L_i)
\right].
\end{equation}
Thus, $\widehat{C}_{95}$ is the fraction of items for which the
panel-matched ensemble CI contains the corresponding subjective MOS\footnote{The indicator function $\mathbb{I}[\cdot]$ equals $1$ if its argument is true and $0$ otherwise.}. Values
closer to the nominal level of $0.95$ indicate better coverage.

\textbf{Listening-Test-Style Significant Differences: }
For each experiment, listener scores are first averaged across items for
every listener--condition pair. A one-factor repeated-measures ANOVA with
factor \emph{condition} is then computed, followed by paired $t$-tests for
all condition pairs with Bonferroni correction at $\alpha=0.05$.

The same procedure is applied to the ensemble: scores are averaged across
items for every ensemble-member--condition pair. Because the number of
ensemble members can differ from the number of listeners, ensemble members
are bootstrapped with replacement to the subjective panel size. A pair is
regarded as objectively significant when at least half of the bootstrap
repetitions yield a Bonferroni-corrected significant difference.

Let $h_p\in\{-1,0,+1\}$ denote the subjective result for pair $p$, where
$0$ denotes a non-significant difference and the sign denotes the preferred
condition. Let $\hat{h}_p$ be the corresponding objective decision. We
report directional significant-pair recall and false-positive rate,

\begin{align}
R_{\mathrm{sig}}
&=
\frac{
\sum_p
\mathbb{I}[h_p\neq0 \land \hat{h}_p=h_p]
}{
\sum_p
\mathbb{I}[h_p\neq0]
},\\
\mathrm{FPR}
&=
\frac{
\sum_p
\mathbb{I}[h_p=0 \land \hat{h}_p\neq0]
}{
\sum_p
\mathbb{I}[h_p=0]
}.
\end{align}
High \(R_{\mathrm{sig}}\) indicates that subjectively significant
differences are detected with the correct direction, whereas low FPR indicates
that the system rarely reports a difference for a subjectively non-significant
pair. These metrics must be interpreted jointly: reporting more pairs as
significantly different may increase recall but can also increase the
false-positive rate.

\section{Results and Discussion}

\subsection{Accuracy, Coverage, and Distributional Agreement}

\label{sec:results_scalar}

\begin{figure}[t]
  \centering
  \newcommand{\imgwidth}{0.495\columnwidth}
  \newcommand{\imgheight}{0.13\textheight}

  \begin{subfigure}[t]{\imgwidth}
    \centering
    \includegraphics[height=\imgheight,
    trim=3mm 1mm 0mm 0mm, clip]{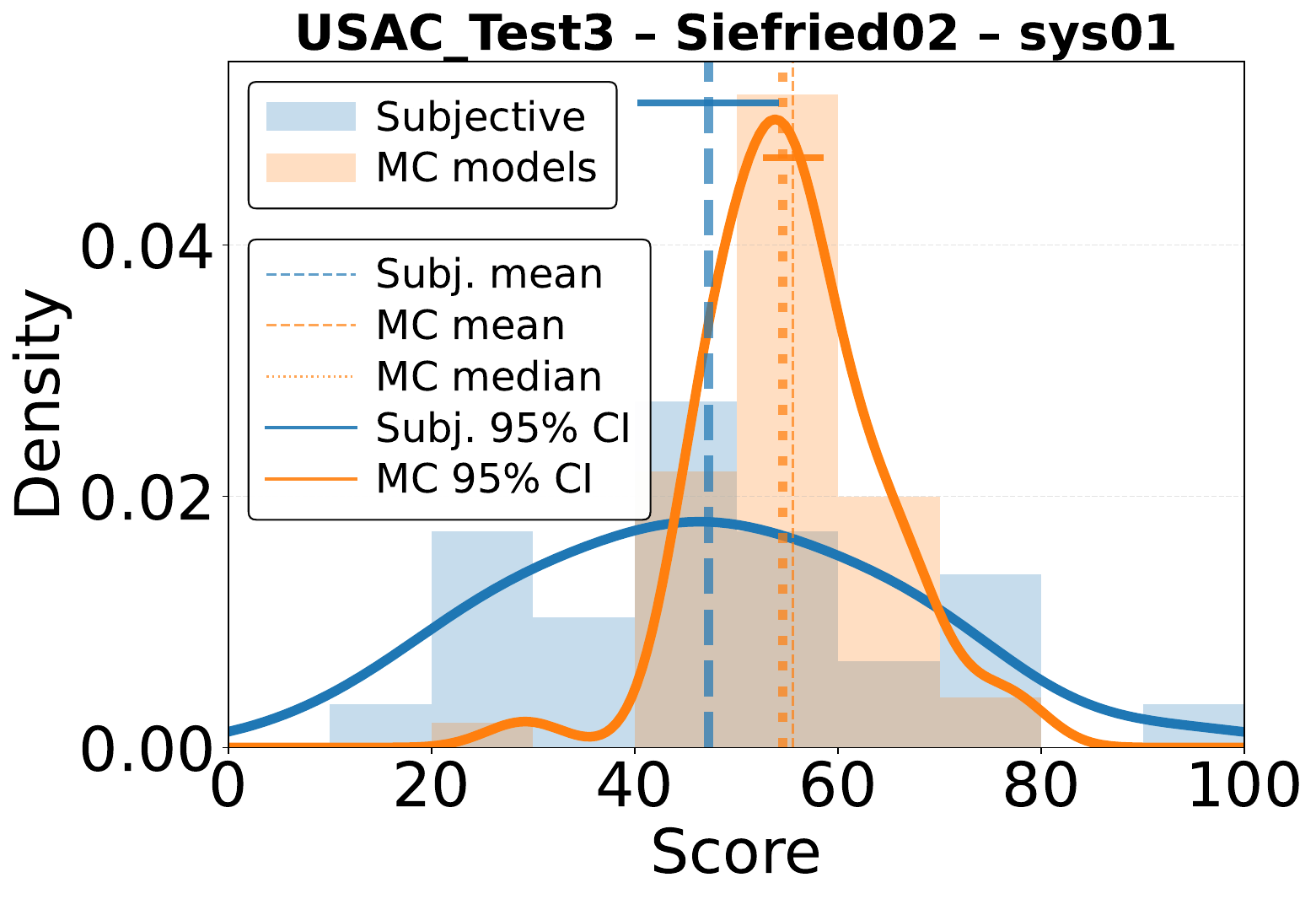}
    \label{fig:hist_before}
  \end{subfigure}
  \hfill
  \begin{subfigure}[t]{\imgwidth}
    \centering
    \includegraphics[height=\imgheight, trim=15mm 1mm 0mm 0mm, clip]{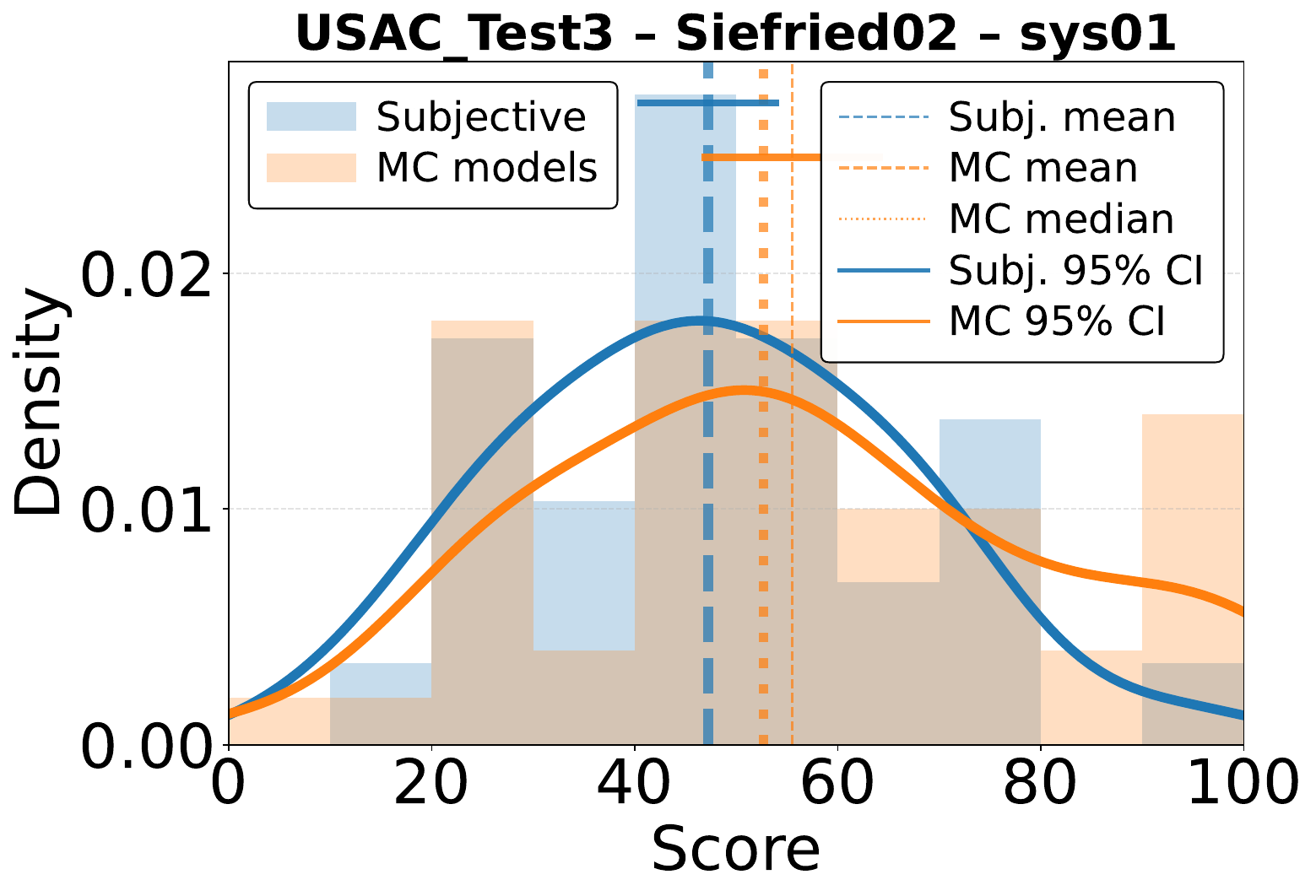}
    \label{fig:hist_after}
  \end{subfigure}
  \caption{Example outputs of ensemble score distributions and corresponding subjective MOS distribution for one item in the evaluation database. Left: before calibration ($D_{KS}=0.48$) and right: after affine mean alignment and width adjustment ($D_{KS}=0.21$).}
  \label{fig:before_after}
\end{figure}

Tab.~\ref{tab:combined_results} reports point estimate MOS accuracy, empirical CI
coverage, distributional agreement, and significance-decision agreement.
Calibration improves coverage on every dataset: from \(0.28\)--\(0.31\) to
\(0.69\)--\(0.82\) on the SEBASS-DB subsets, from \(0.34\) to \(0.47\) on
ODAQ, and from \(0.14\)--\(0.22\) to \(0.33\)--\(0.43\) on USAC. It also
reduces \(\overline{D}_{\mathrm{KS}}\) throughout, indicating better agreement
between the predicted and subjective score distributions (see, e.g.,
Fig.~\ref{fig:before_after}). Coverage nevertheless remains below the nominal
\(0.95\), particularly on USAC. This may partly reflect our deliberate use of
the publicly available SEBASS-DB for reproducible training, at the expense of
a mismatch with the codec, bitrate, and content conditions represented in
USAC.

Affine calibration has little effect on point estimate MOS accuracy; the small PCC
and RMSE changes result from clipping member scores to the MUSHRA scale.
Compared with 2f, the ensemble mean achieves higher PCC and lower RMSE on
every non-aggregated dataset except USAC3, while 2f has a modest advantage on
pooled ODAQ. Unlike the point estimate-only 2f model, the ensemble also provides
item-specific score distributions and panel-size-matched CIs. On the common
USAC1--3 sets, its calibrated PCC/RMSE values are \(0.93/5.67\),
\(0.80/8.39\), and \(0.77/9.52\), whereas GML with CutMix reports PCCs of
\(0.88\), \(0.89\), and \(0.92\), respectively
\cite{jiang2023generative}. Thus, although our method is competitive on
USAC1, it is weaker than GML on USAC2 and USAC3. Nevertheless, these results
are promising given that our lightweight approach uses only 11 fixed PEAQ
features and low-capacity MARS regressors trained on the public SEBASS-DB (11 000 ratings),
whereas GML uses 67\,505 individual ratings from an internal, codec-oriented
corpus and a 15.25-million-parameter network trained on an NVIDIA A100 GPU.
The performance achieved despite the mismatch between SEBASS-DB and the
codec, bitrate, and content conditions in USAC suggests that training on
sufficiently representative public data could further improve
generalization.

\subsection{Pairwise Decision Agreement}
\label{sec:results_decisions}

Calibration makes the paired tests more conservative: FPR decreases or
remains unchanged, but significant-pair recall decreases on every dataset.
These rates can be unstable when based on few pairs: SAOC-DB has only 16
subjectively significant pairs, while USAC1--3 have only two, one, and one
non-significant pairs, respectively.

This trade-off follows from the width adjustment. For conditions \(A\) and
\(B\), calibration preserves their mean contrast while scaling the centered
member variation in paired differences by \(c_{\mathrm{std}}\). For the
complete, unresampled ensemble and without clipping,
\(t_{\mathrm{cal}}=t_{\mathrm{aff}}/c_{\mathrm{std}}\), where
\(t_{\mathrm{aff}}\) denotes the statistic after affine mean calibration.
Because the fitted \(c_{\mathrm{std}}>1\), significance becomes less likely.
Matching marginal score distributions is therefore insufficient for paired
decisions, which also depend on cross-condition covariance.

Reliable listening-test inference would require separate calibration of
marginal variability and cross-condition covariance using listener-level
data. Artificial listener panels should retain the current panel-size matching while also
preserving within-listener dependencies and the target test design. Calibration and decision thresholds should
then be validated on independent listening tests for CI coverage,
false-positive control, and statistical power.

Although the ensemble cannot yet replace panel-level hypothesis testing, it
provides an uncertainty-aware OAQM: its mean estimates MOS, while its
item-dependent spread and approximate CIs help identify predictions requiring
subjective validation.

\section{Summary and Conclusions}
\label{sec:conclusions}

We presented a lightweight, distribution-aware OAQM that produces empirical
score distributions and panel-size-matched CIs in addition to MOS estimates.
Mean-and-width calibration improves CI coverage and distributional agreement
on every evaluated dataset while maintaining competitive MOS accuracy.
Unlike point-estimate metrics, the method provides item-dependent uncertainty;
unlike deep generative listener models, it relies on only 11 PEAQ features,
low-capacity regressors, and exclusively public training data. We therefore
consider ensemble regressors a practical and valuable approach for
uncertainty-aware quality assessment, with clear potential for improvement
through more diverse training data and refined calibration. However,
calibrated marginal distributions do not ensure reliable paired significance
decisions because cross-condition dependencies are not calibrated. The system
is therefore useful for uncertainty screening and prioritising listening
tests, but cannot yet replace panel-level inference. Future work will investigate listener-level calibration of cross-condition
covariance and condition-dependent uncertainty using independent listening
tests. Incorporating out-of-domain detection may further enable the system to
defer automatically when its uncertainty estimates are unlikely to transfer.

\bibliographystyle{IEEEbib}
\bibliography{paper_references}

\end{document}

%% file: method_diagram.tex

\begin{tikzpicture}[
  font=\small,
  >=Latex,
  node distance=3mm,
  every node/.style={
    outer sep=0pt
  },
  signal/.style={
    draw,
    rounded corners,
    align=center,
    inner sep=2.5pt,
    minimum height=10mm,
    fill=gray!10
  },
  process/.style={
    draw,
    rounded corners,
    align=center,
    inner sep=3pt,
    minimum height=13mm,
    fill=blue!7
  },
  ensemblebox/.style={
    draw,
    dashed,
    rounded corners,
    align=center,
    inner sep=3pt,
    minimum height=16mm,
    fill=orange!7
  },
  output/.style={
    draw,
    rounded corners,
    align=center,
    inner sep=3pt,
    minimum height=12mm,
    fill=green!10
  },
  auxiliary/.style={
    draw,
    rounded corners,
    align=center,
    inner sep=2.5pt,
    fill=gray!8,
    font=\footnotesize
  },
  arrow/.style={
    ->,
    thick
  },
  trainarrow/.style={
    ->,
    thick,
    dashed,
    gray!75
  },
  arrowlabel/.style={
    fill=white,
    inner sep=1.5pt,
    align=center,
    font=\footnotesize
  }
]


\node[
  signal,
  text width=19mm
] (ref) {
  Reference\\
  $\mathbf{x}^{\mathrm{ref}}$
};

\node[
  signal,
  text width=19mm,
  below=3mm of ref
] (sut) {
  Test signal\\
  $\mathbf{x}^{\mathrm{sut}}$
};


\node[
  process,
  text width=27mm,
  right=4mm of $(ref.east)!0.5!(sut.east)$
] (peaq) {
  PEAQ-style\\
  perceptual\\
  comparison $F_{PEAQ}$
};

\coordinate (peaq-ref-in) at ($(peaq.west)+(0,3.5mm)$);
\coordinate (peaq-sut-in) at ($(peaq.west)+(0,-3.5mm)$);

\draw[arrow]
  (ref.east) -- ++(2mm,0) |- (peaq-ref-in);

\draw[arrow]
  (sut.east) -- ++(2mm,0) |- (peaq-sut-in);


\node[
  signal,
  text width=25mm,
  right=3mm of peaq
] (features) {
  11 distortion\\
  metrics (MOVs)\\[2mm]
  $\mathbf z\in\mathbb R^{11}$
};

\draw[arrow]
  (peaq) -- (features);


\node[
  ensemblebox,
  text width=34mm,
  right=3mm of features
] (ensemble) {
  \textbf{Bagging ensemble}\\[1mm]
  $f_1\quad f_2\quad\cdots\quad f_K$\\[-1mm]
  {\footnotesize low-capacity regressors}
};

\draw[arrow]
  (features) -- (ensemble);


\node[
  process,
  text width=36mm,
  right=3mm of ensemble
] (calibration) {
  Global calibration $\mathcal C$\\[1mm]
  $s'_k=as_k+b$\\[-1mm]
  $\tilde{s}_k
    =\mu'+c_{\mathrm{std}}(s'_k-\mu')$
};

\draw[arrow]
  (ensemble) --
  node[arrowlabel, above, yshift=1cm] {
    $s_k=f_k(\mathbf z)$
  }
  (calibration);


\node[
  signal,
  text width=30mm,
  right=4mm of calibration
] (sampledist) {
  Calibrated MC\\
  sample distribution\\[1mm]
  $\{\tilde{s}_k\}_{k=1}^{K}$
};

\draw[arrow]
  (calibration) -- (sampledist);

\node[
  output,
  text width=30mm,
  right=4mm of sampledist,
  yshift=8mm
] (mos) {
  MOS estimate\\
  $\displaystyle
   \hat m=\frac{1}{K}
   \sum_{k=1}^{K}\tilde{s}_k$
};

\node[
  output,
  text width=30mm,
  right=4mm of sampledist,
  yshift=-8mm
] (ci) {
  Panel-matched\\
  95\% CI\\[-1mm]
  {\footnotesize $B=4000$, panel size $L$}
};

\coordinate (output-split)
  at ($(sampledist.east)+(2mm,0)$);

\draw[thick]
  (sampledist.east) -- (output-split);

\draw[arrow]
  (output-split) |- (mos.west);

\draw[arrow]
  (output-split) |- (ci.west);


\node[
  auxiliary,
  text width=31mm,
  below=12mm of ensemble
] (training) {
  Training data\\
  $\mathcal D=\{(\mathbf z_i,m_i)\}_{i=1}^{N}$
};

\node[
  auxiliary,
  text width=31mm,
  below=12mm of calibration
] (subjective) {
  Training data statistics\\
  $\mu_i^{\mathrm{subj}},
   \sigma_i^{\mathrm{subj}}$
};

\draw[trainarrow]
  (training.north) --
  node[
    right,
    font=\small,
    text=black!75,
    align=left
  ] {
    bootstrap\\
    samples
  }
  (ensemble.south);

\draw[trainarrow]
  (subjective.north) --
  node[
    right,
    font=\small,
    text=black!95,
    align=left
  ] {
    fit calibration\\
    parameters once
  }
  (calibration.south);

\end{tikzpicture}